\documentclass[letterpaper]{article}

\usepackage{emulateapj}
\usepackage{onecolfloat}
\usepackage{apjfonts}
\usepackage{amsmath}
\usepackage{graphicx}

\begin{document}

\submitted{To Appear in the Astrophysical Journal}

\twocolumn[
\title{X-Ray Spectral and Timing Evolution During the Decay of the
1998 Outburst from the Recurrent X-Ray Transient 4U~1630--47}

\authoremail{jtomsick@ucsd.edu}

\author{John A. Tomsick}
\affil{Department of Physics and Columbia Astrophysics Laboratory, 
Columbia Univ., 550 West 120th Street, New York, NY 10027\nl
(Current address: Center for Astrophysics and Space Sciences, 
Univ. of California, San Diego, MS 0424, La Jolla, CA 92093)\nl
(e-mail: jtomsick@ucsd.edu)}

\vspace{0.1cm}

\author{Philip Kaaret}
\affil{Harvard-Smithsonian Center for Astrophysics, 60 Garden Street, 
Cambridge, MA 02138 (e-mail: pkaaret@cfa.harvard.edu)}

\begin{abstract}

We report on the X-ray spectral and timing behavior of the recurrent
X-ray transient 4U~1630--47 for 51 \it RXTE \rm observations made
during the decay of its 1998 outburst.  The observations began when the
source was still relatively bright, and, during one of the early 
observations, a QPO with a non-Lorentzian profile occurred near 6~Hz.  
As the source decayed, the X-ray flux dropped exponentially with 
an e-folding time of 14.4~d.  The exponential decay was interrupted 
by an increase in the X-ray flux, and a secondary maximum occurred 89~d 
after the onset of the outburst.  A transition marked by significant 
changes in the timing and spectral properties of the source occurred 
104~d after the start of the outburst.  The transition is similar to 
soft-to-hard state transitions observed in other black hole candidate 
X-ray binaries.  Most of the changes associated with the transition 
occurred in less than 2~d.  The timing changes include an increase in 
the continuum noise level from less than 4\% RMS to greater than 10\% RMS 
and the appearance of a quasi-periodic oscillation (QPO) at 3.4~Hz with
an RMS amplitude of 7.3\% in the 2-21~keV energy band.  At the transition, 
the energy spectrum also changed with an abrupt drop in the soft 
component flux in the \it RXTE \rm band pass.  A change in the power-law 
photon index from 2.3 to 1.8, also associated with the transition, 
occurred over a time period of 8~d.  After the transition, the source flux 
continued to decrease, and the QPO frequency decayed gradually from 3.4~Hz 
to about 0.2~Hz.  

\end{abstract}


] 

\section{Introduction}

Although the recurrent X-ray transient and black hole candidate (BHC) 4U~1630--47 has 
been studied extensively since its first detected outburst in 1969 (\cite{priedhorsky86}), 
interest in this source has intensified due to observations made during its 
1998 outburst.  During the 1998 outburst, radio emission was detected for the first 
time (\cite{hjellming99}).  Although the source was not resolved in the radio, the 
optically thin radio emission suggests the presence of a radio jet.  Also, low 
frequency quasi-periodic oscillations (QPOs) were discovered during the 1998 outburst 
(\cite{dieters98a}) using the \it Rossi X-ray Timing Explorer \rm (\it RXTE\rm).

Here, we report on X-ray observations of 4U~1630--47 made with \it RXTE \rm 
(\cite{brs93}) during the decay of its 1998 outburst.  We compare the X-ray light 
curve to those of other BHC X-ray transients and study the evolution of the spectral 
and timing properties during the decay.  Like many other X-ray transients, the light 
curve of 4U~1630--47 shows an exponential decay and a secondary maximum (\cite{csl97}).
During the early part of the decay, when the X-ray flux was high, 4U~1630--47
showed canonical soft state characteristics (\cite{v95};~\cite{nowak95};~\cite{ct96}),
including an energy spectrum with a strong soft component and a steep power-law and 
relatively low timing variability with a fractional RMS (Root-Mean-Square) amplitude 
of a few percent.  Later in the decay, we observe a transition to a spectrally harder 
and more variable state, which has similarities to transitions observed for GS~1124--68 
(\cite{ebisawa94};~\cite{miyamoto94}) and GRO~J1655--40 (\cite{mendez98}) near the 
ends of their outbursts.

In this paper, we describe the 4U~1630--47 X-ray light curve for the 1998 outburst 
and the \it RXTE \rm observations (\S 2).  In \S 3 and \S 4, we present results of 
modeling the power and energy spectra, respectively.  In \S 5, we examine the 
transition in more detail, and \S 6 contains a discussion of the results.  Finally, 
\S 7 contains a summary of our findings.

\section{Observations and Light Curve}

\begin{figure}
\plotone{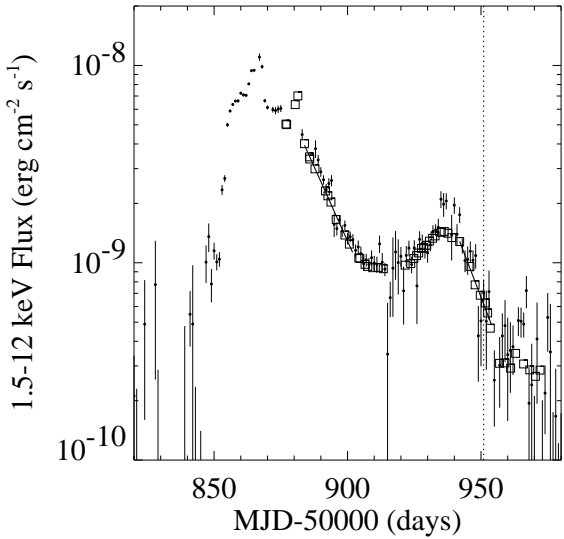}
\caption{X-ray light curve for the 1998 outburst of 4U~1630--47.  
The points with error bars are 1.5-12~keV ASM daily flux measurements,
and the squares mark the 1.5-12~keV fluxes measured by the PCA 
during 51 pointed observations.  The solid lines are exponential
fits to the light curve, and the dotted vertical line marks a transition 
in the spectral and timing properties of the source.\label{fig:lightcurve}}
\end{figure}

We analyzed PCA (Proportional Counter Array) and HEXTE (High Energy X-ray 
Timing Experiment)  data from 51 \it RXTE \rm pointings of 4U~1630--47 during the 
decay of its 1998 outburst.  The observation times, integration times and background 
subtracted 2.5-20~keV PCA count rates are given in Table~\ref{tab:obs}.  In 
Figure~\ref{fig:lightcurve}, we show the 1.5-12~keV PCA fluxes with the ASM 
(All-Sky Monitor) flux measurements in the same energy band.  The ASM light curve was 
produced from data provided by the ASM/\it RXTE \rm teams at MIT and at the \it RXTE \rm 
SOF and GOF at NASA's GSFC.  The 1998 outburst was first detected by BATSE on Modified 
Julian Date 50841 (MJD = JD--2400000.5), and 4U~1630--47 was not detected by the ASM until 
about MJD 50847 (\cite{hjellming99};~\cite{kuulkers98a}).  Figure~\ref{fig:lightcurve} 
shows that the ASM flux increased rapidly after MJD 50850, peaking at 
$1.10\times 10^{-8}$~erg~cm$^{-2}$~s$^{-1}$ (1.5-12~keV) on MJD 50867.  The flux 
dropped to about $6\times 10^{-9}$~erg~cm$^{-2}$~s$^{-1}$ soon after the peak, and 
our \it RXTE \rm observations began during this time.  Our observations fill a gap in the 
ASM light curve near MJD 50880, showing that a flare occurred during this time.  The flux 
decayed exponentially between MJD 50883 and MJD 50902 with an e-folding time of 14.4~d.  
After the exponential decay, the flux increased by about 50\% over a time period of about 
20~d, and a secondary maximum occurred near MJD 50936.  After the secondary maximum, 
the flux decay is consistent with an exponential with an e-folding time of 12~d to 13~d.  
In Figure~\ref{fig:lightcurve}, the vertical dashed line at MJD 50951 marks an 
abrupt change in the timing properties of the source, which is described in detail below.  
The source flux at the transition was between 6 and 
7~$\times 10^{-10}$~erg~cm$^{-2}$~s$^{-1}$.

Soft $\gamma$-ray bursts were detected from a position near 4U~1630--47 on MJD 50979 
(\cite{kouveliotou98}), 7~d after our last \it RXTE \rm observation, and the $\gamma$-ray 
source has been named SGR~1627--41.  Although the position of SGR~1627--41 is not consistent 
with the position of 4U~1630--47 (\cite{hurley99}), the two sources are close enough so that 
they were both in the \it RXTE \rm field of view during our observations, allowing for the 
possibility of source confusion. As described in detail in the appendix, \it RXTE \rm scans 
and \it BeppoSAX \rm observations provide information about possible source confusion.
Based on the evidence, we conclude that it is very unlikely that SGR~1627--41 contributed 
significantly to the flux detected during our observations of 4U~1630--47.

\section{X-Ray Timing}

For each observation, we produced 0.0156-128~Hz power spectra to study the timing 
properties of the system.  For each 64~s interval, we made an RMS normalized
power spectrum using data in the 2-21~keV energy band.  To convert from the Leahy 
normalization (\cite{leahy83}) to RMS, we determined the Poisson noise level
using the method described in Zhang et al.~(1995) with a deadtime of 10 microseconds
per event.  For each observation, the individual 64~s power spectra were averaged,
and the average spectrum was fitted using a least-squares technique and several different 
analytic models.  For individual 64~s power spectra, we calculated the error bars using 
equation A11 from Leahy et al.~(1983).  When combining the power spectra for an entire 
observation, we used two different methods to calculate the errors.  In one method, 
we calculated the errors by propagating the error bars for individual power spectra.
This method does not account for any intrinsic (i.e., non-random) changes in the power 
spectrum over the duration of the observation.  We also estimated the error by calculating 
$\sigma/\sqrt{N}$, where $\sigma$ is the standard deviation of the power measurements from 
the individual spectra, and $N$ is the number of 64~s power spectra being combined.  For 
all observations, the error estimates are approximately the same above $\sim$2~Hz, indicating 
that the shape of the power spectrum at higher frequencies does not change significantly 
during an observation.  However, below $\sim$2~Hz, the calculated errors are significantly
larger using the second method, indicating that intrinsic changes in this region of 
the power spectrum are significant.  In the following, we have used the second method 
to calculate the errors.

\begin{figure}
\plotone{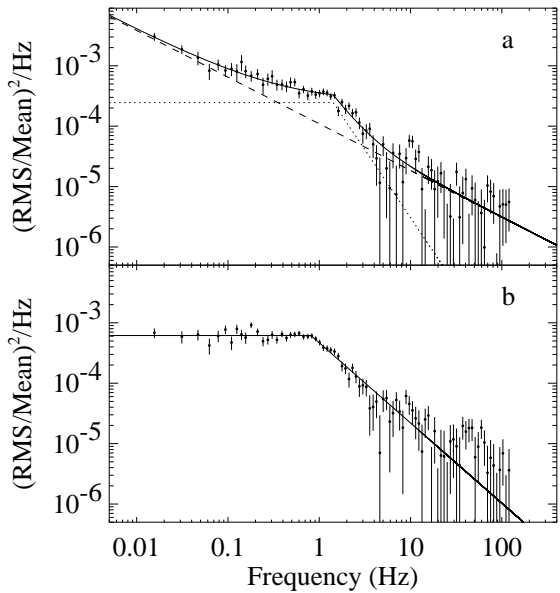}
\vspace{0.7cm}
\caption{The power spectra for observation 8 (a) and observations 11 to 
40 (b).  The observation 8 power spectrum is fitted with a model consisting 
of a power-law (dashed line) and a band-limited (or flat-top) noise component 
(dotted line).  The solid line is the sum of the two components.  For 
observations 11 to 40, only the flat-top component (solid line) is necessary.  
These power spectra have been rebinned for presentation.
\label{fig:powercont}}
\end{figure}

To determine the analytic model to use for the continuum noise, we began by
fitting the power spectrum for each observation with a power-law model.  For some 
observations, the power-law fits are acceptable ($\chi^{2}_{\nu} \sim 1.0$); however,
in most cases, the reduced $\chi^{2}$ is significantly greater than 1.0 and
systematic features appear in the residuals.  Strong QPOs dominate the residuals 
for several observations, and these are discussed in detail below.  For the
observations without obvious QPOs, the power-law residuals are similar and show a 
broad excess peaking between 0.5 and 1.0~Hz.  To model this broad excess, we focus on 
the observation 8 power spectrum since the statistics are good for this observation and 
there are no strong QPOs.  Fitting the observation 8 power spectrum with a power-law 
alone gives a poor fit ($\chi^{2}/{\nu} = 680/444$).  Previous studies of the power 
spectra of BHCs show that the continuum noise can be described by a model consisting of 
two components:  A power-law and a band-limited noise component 
(e.g., Cui et al.~1997;~Miyamoto et al.~1994).  In applying this model to 4U~1630--47, 
we used a broken power-law with the lower power-law index fixed to zero for the 
band-limited component, and hereafter this model is referred to as the flat-top model.  
Applying this two-component model to the observation 8 power spectrum gives a 
significantly improved fit ($\chi^{2}/{\nu} = 486/441$).  Figure~\ref{fig:powercont}a 
shows the observation 8 power spectrum fitted with the two-component model.

\begin{figure}
\plotone{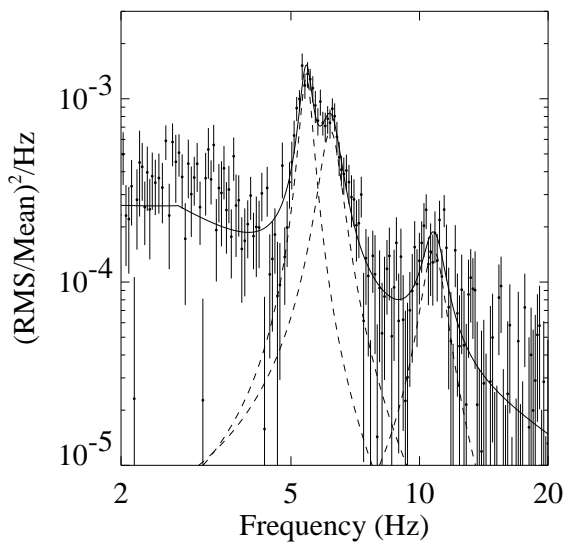}
\caption{Observation 3 power spectrum fitted with the continuum model
(flat-top plus power-law) and three Lorentzians at 5.43~Hz, 6.19~Hz
and 10.79~Hz (dashed lines).  The solid line is the sum of the continuum 
and the Lorentzians.
\label{fig:obs3}}
\end{figure}

For each observation, we fitted the power spectrum using the power-law
model alone, the flat-top model alone and the combination of the two components.  
For several of the observations, the statistics are not good enough to uniquely 
determine the best continuum model.  In these cases, we combined consecutive 
observations, as indicated in Table~\ref{tab:powercont}, to improve the statistics 
and refitted the power spectra with the same models.  For observations 1 to 10, the 
fit using the two-component model is significantly better than using either of the 
individual components, indicating that these power spectra require both components.  
For observations 11 to 51, the flat-top model alone provides a significantly better 
fit than the power-law model alone, and the two-component model does not provide a 
significantly better fit than the flat-top component alone.  We conclude that only 
the flat-top component is necessary to fit these power spectra.  The continuum 
parameters for all observations are given in Table~\ref{tab:powercont}.  In cases 
where the power-law component is not significantly detected, the 90\% confidence 
upper limit on the contribution from a power-law with an index of $-1.0$ is given.  
Figure~\ref{fig:powercont}b shows the power spectrum for observations 11 to 40 
combined, illustrating that the power-law component is not significant at low
frequencies.  We note that there is some evidence for excess noise near 45~Hz, but 
this excess is not statistically significant.  For observations 41 to 51, the RMS 
amplitude for the continuum noise is 10\% to 17\%, which is considerably higher 
than for observations 1 to 40.  In determining the continuum parameters, we 
included Lorentzians to model the QPOs as marked in Table~\ref{tab:powercont}.

To determine where QPOs are present, we examined the residuals for fits with the 
continuum model only.  For observations 1-2, 3, 6, 7, 8, 41, 42, 43, 44, 45, 
46-48 and 49, systematic features in the residuals suggest the presence of QPOs.  
To determine if these features are statistically significant, we compared the 
$\chi^{2}$ for a fit with the continuum model only to a fit with a Lorentzian 
added to the continuum model.  F-tests indicate that QPOs significant at greater 
than 96\% confidence occurred for observations 1-2, 3, 41, 42, 43 and 46-48.  
For observation 1-2, the continuum model provides a relatively poor fit to the 
data ($\chi^{2}/\nu = 561/441$), and the largest residuals occur near 11~Hz.  
The fit is significantly improved ($\chi^{2}/\nu = 471/438$) when a Lorentzian 
is added to the continuum model.  The QPO centroid, FWHM and RMS amplitude are 
$10.8\pm 0.2$~Hz, $2.9\pm 0.6$~Hz and $2.01\%\pm 0.16$\%, respectively.  Although 
the features for observations 6, 7 and 8 are not as statistically significant, 
they also have centroids between 10 and 13~Hz and may be related to the 
observation 1-2 QPO.

For observation 3, the continuum model provides an extremely poor fit
($\chi^{2}/\nu = 987/441$), and the largest residuals occur near 6~Hz.  Although
the fit is significantly improved by the addition of a Lorentzian at 5.7~Hz, the
fit is still relatively poor ($\chi^{2}/\nu = 685/438$), and systematic features
are present in the residuals, which indicate that the 5.7~Hz QPO is not
well-described by a Lorentzian.  As for some other BHCs 
(\cite{belloni97};~\cite{rtb99}), the QPO has a high frequency shoulder
that can be modeled using a second Lorentzian.  Modeling the QPO with
Lorentzians at 5.4~Hz and 6.2~Hz improves the fit to $\chi^{2}/\nu = 608/435$.
The fit can be further improved to $\chi^{2}/\nu = 552/432$ by the addition of 
a QPO near 11~Hz.  It is possible that the 11~Hz QPO is a harmonic of the lower
frequency QPO, but it may also be related to the QPO that occurred during 
observation 1-2.  Table~\ref{tab:obs3} summarizes the QPO parameters for 
observation 3, and it should be noted that three Lorentzians were included in 
the model in determining the continuum parameters given in 
Table~\ref{tab:powercont}.  Figure~\ref{fig:obs3} shows the observation 3 power 
spectrum fitted with a model consisting of the continuum plus three Lorentzians 
to model the QPOs.  The Lorentzians at $5.43\pm 0.02$~Hz, $6.19\pm 0.04$~Hz and 
$10.79\pm 0.14$~Hz have RMS amplitudes of $2.89\%\pm 0.18$\%, $2.85\%\pm 0.21$\% 
and $1.85\%\pm 0.20$\%, respectively.  To determine if the QPO properties 
changed during the observation, we divided observation 3 into two time segments
with durations of 576~s and 512~s, made power spectra for each segment and 
fitted the power spectra with a model consisting of the continuum (flat-top 
plus power-law) plus three Lorentzians.  The results for these fits are given
in Table~\ref{tab:obs3}.  There is no evidence for large changes in the QPO
properties between the two time segments.  

\begin{figure}
\plotone{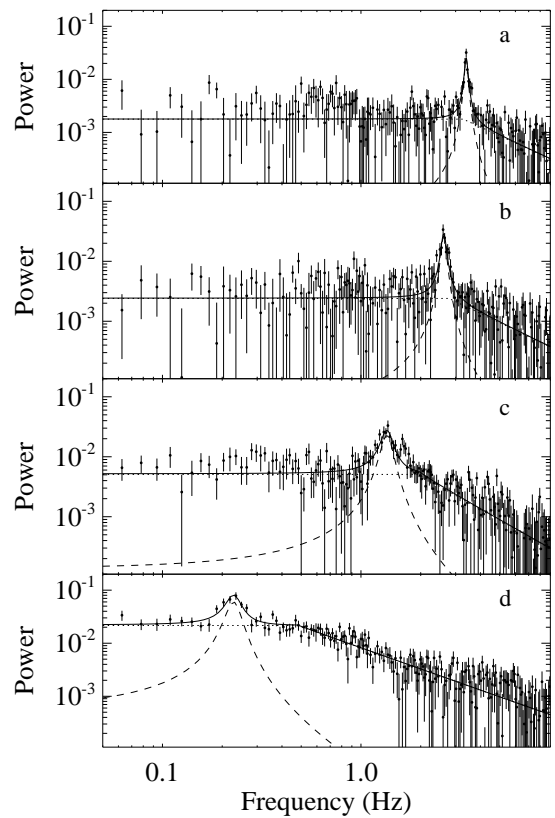}
\vspace{1.0cm}
\caption{Power spectra (Power = (RMS/Mean)$^{2}$/Hz) for observations 
41 (a), 42 (b), 43 (c) and 46-48 (d) showing QPOs detected at 3.390~Hz, 
2.613~Hz, 1.351~Hz and 0.228~Hz.  The power spectra are fitted with a model 
consisting of a flat-top (dotted line) and a Lorentzian (dashed line).  The 
solid line is the sum of the two components.
\label{fig:shiftqpo}}
\end{figure}

The increase in the continuum noise level that occurred between observations 
40 and 41 was accompanied by the appearance of a QPO at $3.390\pm 0.008$~Hz with 
an RMS amplitude of $7.30\%\pm 0.33$\%.  In subsequent observations, the QPO 
frequency gradually shifted to lower frequency.  Figure~\ref{fig:shiftqpo} shows 
the power spectra for observations 41, 42, 43 and 46-48.  After the 3.4~Hz QPO 
appeared for observation 41, QPOs occurred at $2.613\pm 0.012$~Hz, 
$1.351\pm 0.012$~Hz and $0.228\pm 0.003$~Hz for observations 42, 43 and 46-48, 
respectively.  We note that the observation 43 QPO shows some evidence 
for a high frequency shoulder.  QPOs with lower statistical significance occurred 
for observations 44, 45 and 49 with frequencies of $0.430\pm 0.006$~Hz, 
$0.365\pm 0.011$~Hz and $0.182\pm 0.005$~Hz.  It should be noted that these QPOs 
are consistent with the gradual shift to lower frequencies.  The QPO parameters 
for observations 41 to 49 are given in Table~\ref{tab:shiftqpo}.  

\section{Energy Spectra}

We produced PCA and HEXTE energy spectra for each observation using the processing
methods described in Tomsick et al.~(1999).  We used the PCA in the 2.5-20~keV
energy band and HEXTE in the 20-200~keV energy band.  For the PCA, we used standard 
mode data, consisting of 129-bin spectra with 16~s time resolution, included 
only the photons from the top anode layers and estimated the background using the 
sky-VLE model\footnote{%
See M.J. Stark et al.~1999, PCABACKEST, available at 
http:// lheawww.gsfc.nasa.gov/docs/xray/xte/pca.}.  We used the version 2.2.1 response 
matrices with a resolution parameter of 0.8 and added 1\% systematic errors to account 
for uncertainties in the PCA response.  As described in Tomsick et al.~(1999), we used
Crab spectra to test the response matrices and found that the response matrix 
calibration is better for PCUs 1 and 4 than for the other three Proportional
Counter Units (PCUs); thus, we only used these two PCUs for spectral analysis and 
allowed for free normalizations between PCUs.  PCU 4 was off during three observations
(34, 39 and 48), and, to avoid instrumental differences, we did not use these 
observations in our spectral analysis.  Previously, we found that the PCA over-estimates 
the source flux by a factor of 1.18 (\cite{tomsick99}), and, in this paper, we
reduced the fluxes and spectral component normalizations by a factor of 1.18 so that 
the PCA flux scale is in agreement with previous instruments.  

HEXTE energy spectra were produced using standard mode data, consisting of 64-bin 
spectra with 16~s time resolution.  We used the March~20,~1997 HEXTE response matrices
and applied the necessary deadtime correction (\cite{rothschild98}).  For the spectral 
fits, the normalizations were left free between cluster A and cluster B.  It is 
well-known that the HEXTE and PCA normalizations do not agree, so the normalizations
were left free between HEXTE and the PCA.  The HEXTE background subtraction is
performed by rocking on and off source.  Each cluster has two background fields, 
and we checked the HEXTE background subtraction by comparing the count rates for
the two fields.  In cases where contamination of one of the fields occurred, we 
only used the data from the non-contaminated background field.

We first fitted the energy spectra using a power-law with interstellar absorption, 
but this model does not provide acceptable fits to any of the spectra.  For most of 
the observations, the residuals suggest the presence of a soft component, which is 
typical for 4U~1630--47 (\cite{tomsick98};~\cite{parmar97}).  A soft component was 
also detected during \it BeppoSAX \rm observations of 4U~1630--47, which overlap 
with our \it RXTE \rm observations (\cite{oosterbroek98}).  Since 
Oosterbroek et al.~(1998) found that a disk-blackbody model (\cite{makishima86}) 
provides a good description of the soft component observed by \it BeppoSAX\rm, we 
added a disk-blackbody model to the power-law component and refitted the \it RXTE \rm
spectra.  Although the addition of a soft component improves the fits significantly
in most cases, the fits are only formally acceptable for a small fraction of the 
observations, and, in the worst case, the reduced $\chi^{2}$ is 3.1 for 106 
degrees of freedom.  

A broad iron absorption edge, associated with the Compton reflection component
(\cite{lw88}), is commonly observed in the energy spectra of BHCs 
(\cite{ebisawa94} and references therein;~\cite{sobczak99}).  
We refitted the 4U~1630--47 spectra with the model given in equation 3 of
Ebisawa et al.~(1994), which includes a broad absorption edge in addition to
the disk-blackbody and power-law components.  Following Ebisawa et al.~(1994),
we fixed the width of the absorption edge to 10~keV and left the edge energy 
free.  For all of the 4U~1630--47 observations, the fits are significantly better 
with the absorption edge.  As an example, for observation 8, the fit improved from 
$\chi^{2}/\nu = 179/106$ using the disk-blackbody plus power-law model without 
the edge to $\chi^{2}/\nu = 110/104$, indicating that the edge is required at
the 99.1\% confidence level.  In addition to the absorption edge, an iron emission 
line is expected due to fluorescence of the X-ray illuminated accretion disk 
material (\cite{matt92}); thus, we have added an emission line to our model to 
determine whether the line is present in the spectra.  We used a narrow 
emission line since the width of the emission line could not be constrained,
and the energy of the emission line was a free parameter.  

\begin{figure}
\plotone{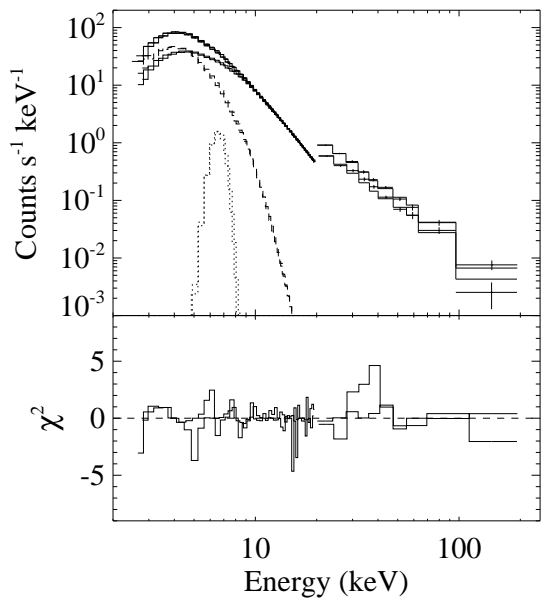}
\vspace{-0.8cm}
\caption{PCA and HEXTE energy spectrum for observation 8
folded with the instrument response and fitted with a model
consisting of a disk-blackbody (dashed line), a power-law
(thin solid line), a narrow emission line (dotted line) and a 
broad iron absorption edge.  The sum of these components is
marked with a thick solid line.  The column density is fixed to 
$9.45\times 10^{22}$~cm$^{-2}$.  The bottom panel shows the 
residuals for the fit.\label{fig:energy}}
\end{figure}

We fitted the spectra with the column density free and also with the column 
density fixed to the mean value for the 51 observations, 
$9.45\times 10^{22}$~cm$^{-2}$.  For all observations, the quality of
the fit is not significantly worse with the column density fixed.
Table~\ref{tab:energy} shows the results for the spectral fits with the column 
density fixed using a model consisting of a power-law, a disk-blackbody component, 
a narrow emission line and a broad absorption edge.  The free parameters for the 
power-law component are the photon index ($\Gamma$) and the normalization.  For the 
disk-blackbody component, the temperature at the inner edge of the disk ($kT_{in}$) 
and the normalization are free parameters.  Rather than the power-law and 
disk-blackbody normalizations, the component fluxes are given in 
Table~\ref{tab:energy}.  The emission line energy ($E_{line}$) and normalization 
($N_{line}$) and the edge energy ($E_{edge}$) and optical depth ($\tau_{\rm{Fe}}$) 
are free parameters.  However, in cases where the best fit value for $E_{edge}$ 
is less than 7.1~keV (the value for neutral iron), we fixed $E_{edge}$ to 7.1~keV.
In Table~\ref{tab:energy}, we do not give error estimates for $kT_{in}$ since the 
uncertainty for this parameter is dominated by systematic error due to uncertainty 
in the correct value for the column density.  By comparing the values found
for $kT_{in}$ with the column density fixed to those with the column density
free, we estimate that the systematic error is 0.05~keV.  For the 51 
observations, the largest $\chi^{2}_{\nu}$ is 1.32 for 102 degrees of freedom 
and $\chi^{2}_{\nu}<1.0$ for 44 of the observations, indicating that the spectra 
are well-described by the model.  Figure~\ref{fig:energy} shows the observation 8 
energy spectrum and residuals.  The residuals shown in Figure~\ref{fig:energy} 
typify the quality obtained for the observations.

\begin{figure}
\plotone{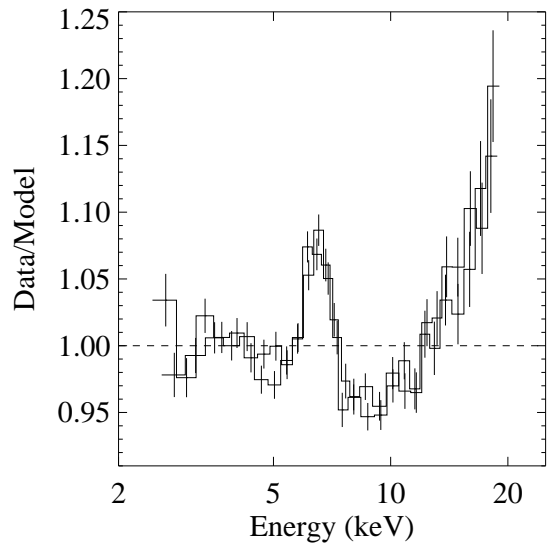}
\caption{Data-to-model ratio for a power-law plus disk-blackbody
fit to the spectrum for observations 44 to 51.  An emission
line at $6.46\pm 0.04$~keV with an equivalent width of 91~eV is 
clearly present.  Although we fitted both the PCA and the HEXTE data, 
only the PCA data are shown.\label{fig:line}}
\end{figure}

For each observation, we determined the significance of the emission line by 
refitting the spectra without the line and using an F-test.  In cases where the 
significance of the emission line is less than 90\%, we fixed $E_{line}$ to the 
best fit value and determined the 90\% confidence upper limit on $N_{line}$.  
Although most of the spectra do not require the emission line at a high confidence 
level, the line is required at greater than 90\% confidence for 16 of the 51 
observations, and at greater than 95\% confidence for 9 observations.  In the 
cases where the iron line is detected at greater than 90\% confidence, the 
equivalent width of the iron line is between 45~eV (for observation 7) and 
110~eV (for observation 47).  

We also determined the significance of the disk-blackbody component using
the same method described above for the emission line.  With the column 
density fixed, the disk-blackbody component is required at greater than
97\% confidence for every observation; however, with the column density
free, the disk-blackbody component is not required for several observations.
With the column density free, the disk-blackbody components are significant
at only 50\% and 65\% confidence for observations 3 and 4, respectively, and 
at between 46\% and 70\% confidence for observations 41 to 51.  In 
Table~\ref{tab:energy}, the disk-blackbody fluxes for these observations are 
marked as upper limits since the component is not detected.  For observations 
41 to 51, the best fit values of $kT_{in}$ are also marked as upper limits 
since the peak of the disk-blackbody flux falls below the PCA band pass and 
we cannot constrain $kT_{in}$ and the column density independently.

The flux levels and line parameters are similar for observations 44 to 51 so 
we refitted the combined spectrum for these observations.  As shown in
Table~\ref{tab:energy}, an emission line at $6.46\pm 0.04$~keV is detected at 
99.93\% confidence.  The line energy is consistent with emission from neutral 
or mildly ionized iron and the line equivalent width is 91~eV.  We also fitted 
the combined spectrum with a model consisting of a disk-blackbody and a 
power-law, and Figure~\ref{fig:line} shows the data-to-model ratio, clearly 
indicating the presence of the iron line.  Since 4U~1630--47 lies 
along the Galactic ridge ($l = 336.91^{\circ}$, $b = 0.25^{\circ}$), we have 
considered the possibility that the 4U~1630--47 spectra are contaminated
by Galactic ridge emission.  It is unlikely that the ridge emission is the
source of the iron line detected in our spectra because the line energy
we observe is considerably lower than the values measured by $ASCA$, 
$Ginga$ and $Tenma$ for the Galactic ridge, which are all near 6.7~keV 
(\cite{kaneda97} and references therein).  Also, based on the spectrum 
of the Galactic ridge emission measured by \it RXTE \rm (\cite{vm98}), 
the spatially averaged Galactic ridge 2.5-20~keV flux is only 6\% 
of the flux for the combination of observations 44 to 51, indicating that 
the level of contamination by the Galactic ridge emission should be low.

\section{State Transition}

\begin{figure}
\plotone{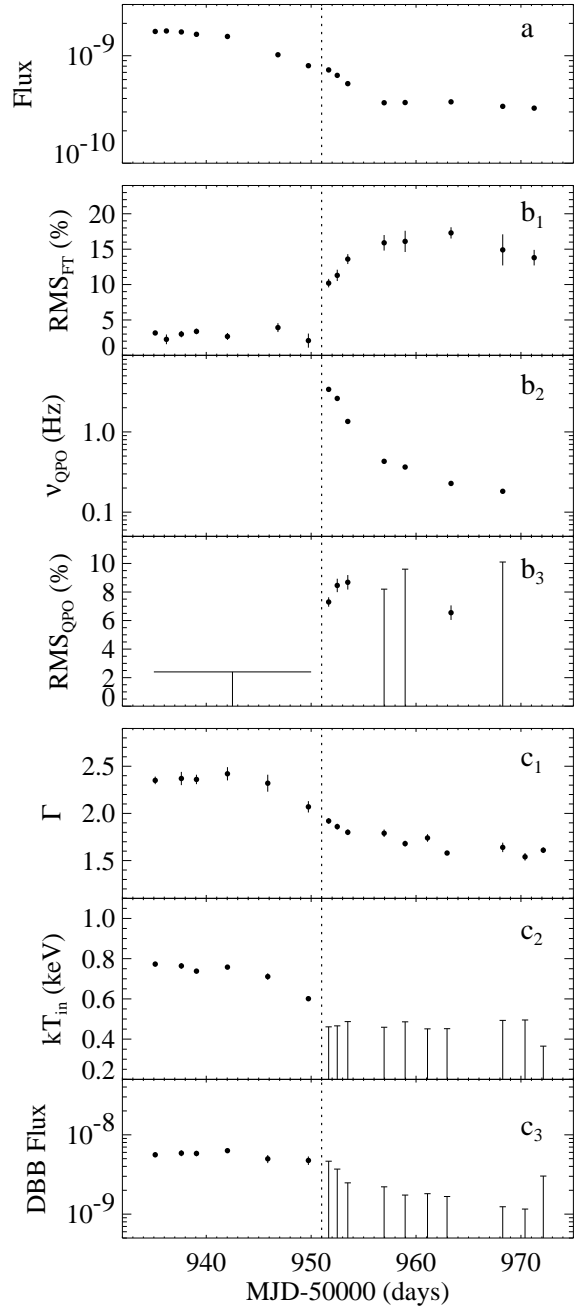}
\vspace{1.0cm}
\caption{Timing and Spectral parameters for observations 33 to 51.  
Panel a is the 2.5-20~keV flux (in ergs~cm$^{-2}$~s$^{-1}$) vs. time.  
The timing parameters are shown in panels b$_{1}$, b$_{2}$ and b$_{3}$, 
and the spectral parameters are shown in panels c$_{1}$, c$_{2}$ and 
c$_{3}$.  The error bars displayed correspond to $\Delta \chi^{2} = 1.0$
(68\% confidence) and the upper limits shown are 90\% confidence.
The bolometric disk-blackbody flux is shown in panel c$_{3}$.  A 
vertical dotted line at MJD 50951 marks the state transition.\label{fig:trans}}
\end{figure}

Figure~\ref{fig:trans} shows the evolution of the timing and spectral 
parameters for observations 33 to 51.  Significant changes in the 4U~1630--47 
emission properties occurred between observations 40 and 41, and we interpret 
this as evidence that a state transition occurred.  In Figure~\ref{fig:trans}, 
the transition is marked with a vertical dashed line at MJD 50951.  At the 
transition, an increase in source variability occurred with the 0.01-10~Hz
RMS amplitude of the flat-top component increasing from between 2.1\% and 3.9\% 
for observations 33 to 40 to $10.2\%\pm 0.6$\% for observation 41.  As shown 
in panel b$_{1}$ of Figure~\ref{fig:trans}, the RMS amplitude continued to 
increase after the transition, reaching a maximum value of $17.3\%\pm 0.8$\% 
for observation 46-48.  In addition to the increase in the continuum noise level, 
a QPO appeared for observation 41, and the centroid QPO frequency and RMS amplitude 
are shown in panels b$_{2}$ and b$_{3}$, respectively.  The timing changes 
occurred in less than 2~d and with only a small change in the 1.5-12~keV flux 
(shown in panel a of Figure~\ref{fig:trans}).  

To determine if a QPO was present before the transition, we made a combined 
power spectrum for observations 33 to 40.  When the 2-21~keV power spectrum is 
fitted with a flat-top model, the residuals show no clear evidence for a QPO.  
The 90\% confidence upper limit on the RMS amplitude for a QPO in a frequency 
range from 0.1~Hz to 10~Hz is 2.4\%.  We performed an additional test by determining 
the energy range where the observation 41 QPO is strongest.  For observation 41,
the RMS amplitudes are $6.1\%\pm 0.4$\% and $8.6\%\pm 0.4$\% for the 2-6~keV and 
6-21~keV energy bands, respectively, indicating that the strength of the QPO 
increases with energy.  Since the QPO is stronger in the 6-21~keV energy band for 
observation 41, we produced a 6-21~keV power spectrum for observations 33 to 40.
As before, when a flat-top model is used to fit the power spectrum, the residuals
do not show evidence for QPOs, and the 90\% confidence upper limit on the RMS 
amplitude for a QPO in a frequency range from 0.1~Hz to 10~Hz is 2.9\%.

\begin{figure}
\plotone{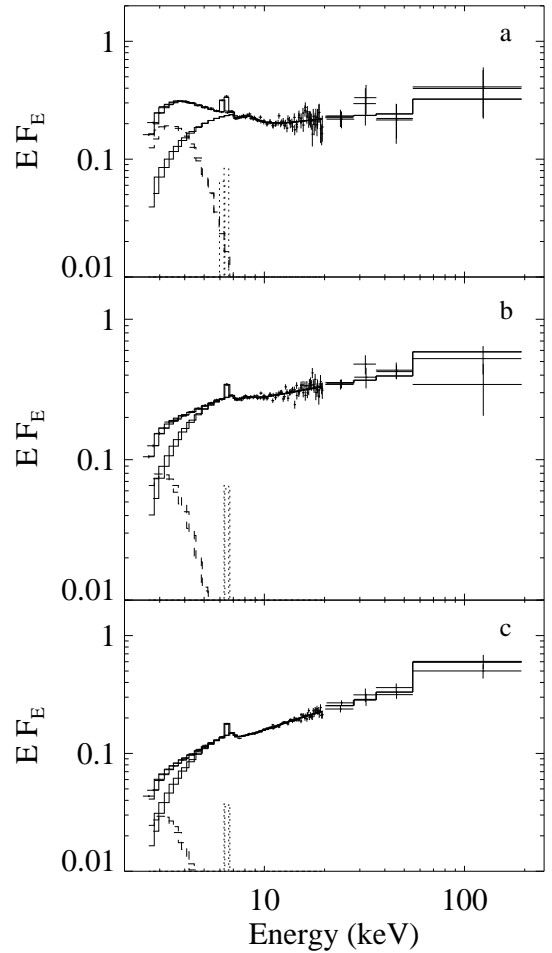}
\vspace{1.0cm}
\caption{Unfolded PCA and HEXTE energy spectra for (a) Observation 40; 
(b) Observation 41; and (c) Observations 44 to 51 with the spectral
components marked as for Figure~\ref{fig:energy}.  The figure shows that 
the disk-blackbody temperature changes significantly between observations 
40 and 41, and that the disk-blackbody flux is significantly lower for 
observations 44 to 51.  Also, the power-law gradually hardens. 
\label{fig:energy3}}
\end{figure}

Although the difference between the observation 40 and 41 energy spectra is 
not as distinct as for the power spectra, changes occurred.  In Figure~\ref{fig:trans}, 
the spectral parameters $\Gamma$ and $kT_{in}$ are shown in panels c$_{1}$ 
and c$_{2}$, respectively.  The power-law index hardened slightly between 
observations 40 and 41; however, this change appears to be part of a larger 
trend, which occurred over a span of 8~d between observations 38 and 43.  
The inner disk temperature began to decrease near observation 37, and 
the soft component is not confidently detected after observation 40, which
probably indicates that $kT_{in}$ continued to drop after observation 40.
The spectral changes are also illustrated in Figures~\ref{fig:energy3}a and 
\ref{fig:energy3}b, which show the energy spectra for observations 40 and 
41, respectively.  Figure~\ref{fig:energy3}c shows the energy spectrum
for observations 44 to 51, indicating that the spectrum continued to
harden after the transition.

In summary, during the transition, the noise level increased, the power-law 
spectral index hardened and the soft component flux in the $RXTE$ band pass 
decreased.  Similar changes are typically observed in BHC systems when 
soft-to-hard state transitions occur (\cite{v95};~\cite{nowak95};~\cite{ct96}),
and we conclude that such a transition occurred for 4U~1630--47.  We also show 
that QPOs were not present during the observations leading up to the transition, 
indicating that their appearance during observation 41 is related to the state 
transition.

\section{Discussion}

\subsection{Comparisons to Previous 4U~1630--47 Outbursts}

Since 4U~1630--47 was discovered in 1969, quasi-periodic outbursts have been observed
from this source every 600 to 690~d (\cite{kuulkers97})\footnote{%
However, the 1999 outburst significantly deviates from this periodicity 
(\cite{mccollough99}).}.  The light curve for the 1998 
4U~1630--47 outburst is the best example of a ``fast-rise exponential-decay" (or FRED) 
light curve (Chen et al.~1997) that has been observed for 4U~1630--47.  A FRED light 
curve may have been observed for 4U~1630--47 by the $Vela~5B$ X-ray monitor in 1974 
(\cite{priedhorsky86};~Chen et al.~1997), but the temporal coverage was sparse compared to 
the coverage obtained for the 1998 outburst.  Good temporal coverage was obtained for 
the 1996 outburst by the $RXTE$/ASM, and a FRED light curve was not observed.  After 
the start of the 1996 outburst, the flux stayed at a high level for about 100~d before 
decaying exponentially with an e-folding time of about 14.9~d (\cite{kuulkers97}).  
Although the overall light curve shapes are different for the two outbursts, it is
interesting that the e-folding time for the 1998 outburst, 14.4~d, is close to the
14.9~d e-folding time for the 1996 outburst.  This may suggest that the e-folding time 
is related to a physical property of the system that does not change between outbursts.
For example, the e-folding time may be related to the mass of the compact object 
(\cite{ccl95}) or the radius of the accretion disk (\cite{kr98}).  

A state transition with similarities to the soft-to-hard transition we report in this 
paper was observed by $EXOSAT$ during the decay of the 1984 outburst from 4U~1630--47.  
Four $EXOSAT$ observations of 4U~1630--47 were made during outburst decay (\cite{psw86}).  
During the first two observations in 1984 April and 1984 May, a strong soft component was 
observed in the energy spectrum.  The power-law was harder in May than in April and became 
even harder for two observations made in 1984 July.  During the July observations, the soft 
component was not clearly detected.  Assuming a soft-to-hard transition occurred between
May and July, the transition took place at a luminosity between $10^{36}$~erg~s$^{-1}$ and 
$10^{38}$~erg~s$^{-1}$ (1-50~keV), which is consistent with the luminosity where the 1998 
soft-to-hard transition occurred, $7\times 10^{36}$~erg~s$^{-1}$ (2.5-20~keV).  The 
luminosities given here are for an assumed distance of 10~kpc; however, the distance to 
4U~1630--47 is not well-determined.

\subsection{Comparisons to Other Black Hole Candidate X-Ray Transients}

Here, we compare the properties 4U~1630--47 displayed during the decay of its 1998 
outburst to those observed for other X-ray transients.  We have compiled a list of 
comparison sources using Tanaka \& Shibazaki~(1996) and Chen, Shrader \& Livio (1997).  
The comparison group contains the BHC X-ray transients that had strong soft 
components during outburst and FRED light curves.  The comparison sources from the 
above references are GS~1124--68, GS~2000+251, A~0620--00, EXO~1846--031, Cen~X-2, 
4U~1543--47 and A~1524--617.  We also include a recent X-ray transient, XTE~J1748--288, 
that has similar properties to this group.  For the eight comparison sources, the 
exponentially decaying portions of their X-ray light curves have e-folding times ranging 
from 15~d to 80~d (Chen et al.~1997;~Revnivtsev et al.~1999), and the mean decay time 
is 39~d.  Thus, the 14.4~d e-folding time for 4U~1630--47 is shorter than average, but 
not unprecedented.

Like 4U~1630--47, secondary maxima occurred in the X-ray light curves of 4U~1543-47, 
A~0620--00, GS~2000+251 and GS~1124--68, and a tertiary maximum occurred for A~0620--00  
(\cite{kaluzienski77}).  It is likely that the secondary and tertiary maxima are the 
result of X-ray irradiation of the outer accretion disk or the optical companion
(\cite{kr98};~\cite{clg93};~\cite{aks93}).  In this picture, the time between the start 
of the outburst and subsequent maxima depends on the viscous time scale of the disk.
For A~0620--00, GS~2000+251 and GS~1124--68, secondary maxima are observed 55 to 75~d 
after the start of the outburst.  These maxima, often referred to as ``glitches", consist 
of a sudden upward shift in X-ray flux, interrupting the exponential decay.  The tertiary 
maximum observed for A~0620--00 about 200~d after the start of the outburst is significantly 
different, and can be described as a broad (35 to 40~d) bump in the X-ray light curve near 
the end of the outburst.  The 4U~1630--47 secondary maximum is similar to the A~0620--00 
tertiary maximum since it is a broad (about 25~d) increase in flux near the end of the 
outburst.  However, the secondary maximum peaked about 89~d after the start of the outburst, 
which is considerably less than for A~0620--00.

Four sources in our comparison group exhibited soft-to-hard state transitions
during outburst decay:  A~0620--00 (\cite{kuulkers98b}), GS~2000+251 (\cite{ts96}), 
GS~1124--68 (\cite{kitamoto92}) and XTE~J1748--288 (Revnivtsev et al.~1999).  The 
4U~1630--47 transition occurred 104~d after the start of the outburst, while 
transitions for the other four sources occurred 100 to 150~d, 230 to 240~d, 
131 to 157~d and about 40~d after the starts of the outbursts for A~0620--00, 
GS~2000+251, GS~1124--68 and XTE~J1748--288, respectively.  Detailed X-ray spectral 
and timing information is available after the transition to the hard state for 
GS~1124--68.  Like 4U~1630--47, the GS~1124--68 transition was marked by an increase 
in the RMS noise amplitude; however, in contrast to 4U~1630--47, QPOs were not observed 
for GS~1124--68 in the hard state (\cite{miyamoto94}).  Also, during the GS~1124--68 
transition, the X-ray spectrum hardened with a drop in the inner disk temperature 
($kT_{in}$) and a change in the power-law photon index ($\Gamma$) from 2.2 to 1.6 
(\cite{ebisawa94}).  During the 4U~1630--47 transition, the change in the soft component 
was consistent with a drop in $kT_{in}$, and $\Gamma$ changed from 2.3 to 1.8.  
While the $Ginga$ observations of GS~1124--68 were relatively sparse near the 
transition, our observations of 4U~1630--47 show that soft-to-hard transitions can 
occur on a time scale of days.

\subsection{Hard State QPOs}

Although QPOs were not detected after the GS~1124--68 state transition, QPOs were observed 
after a similar transition for the microquasar GRO~J1655--40 during outburst decay 
(\cite{mendez98}).  \it RXTE \rm observations of GRO~J1655--40 show that a state transition 
occurred between 1997 August 3 and 1997 August 14.  The transition was marked by an 
increase in the continuum variability from less than 2\% RMS to 15.6\% RMS, a decrease in 
the characteristic temperature of the soft spectral component ($kT_{in}$) from 0.79~keV to 
0.46~keV and the appearance of a QPO at 6.46~Hz with an RMS amplitude of 9.8\%.  A QPO was 
also detected at 0.77~Hz during an August 18 \it RXTE \rm observation of GRO~J1655--40 
when the 2-10~keV flux was about a factor of four lower than on August 14; thus, the shift 
to lower frequencies with decreasing flux is common to GRO~J1655--40 and 4U~1630--47.  The 
correlations between spectral and timing properties for the microquasar GRS~1915+105 are 
similar to those observed for GRO~J1655--40 and 4U~1630--47.  Markwardt, Swank \& Taam 
(1999) and Muno, Morgan \& Remillard (1999) found that 1-15~Hz QPOs are observed for 
GRS~1915+105 more often when the source spectrum is hard.  Markwardt et al.~(1999) report 
a correlation between QPO frequency and disk flux, and Muno et al.~(1999) find that the 
QPO frequency is correlated with $kT_{in}$.  Although these results suggest that the 
QPO is related to the soft component, the fact that the QPO strength increases with 
energy for 4U~1630--47, GRO~J1655--40 and GRS~1915+105 indicates that the QPO mechanism 
modulates the hard component flux. 

A physical model that has been used to explain the energy spectra of 
BHC systems involves the presence of an advection-dominated accretion flow 
or ADAF (\cite{narayan97}).  The model assumes the accretion flow 
consists of two zones:  An optically thin ADAF region between the black 
hole event horizon and a transition radius, $r_{t}$, and a geometrically 
thin, optically thick accretion disk outside $r_{t}$.  Esin, McClintock \& 
Narayan (1997) developed and used this model to explain the spectral changes 
observed for GS~1124--68 during outburst decay, which are similar to the 
spectral changes observed for 4U~1630--47.  The different emission states 
observed during the decay can be reproduced by decreasing the mass accretion 
rate and increasing $r_{t}$.  This model suggests that the gradual decrease 
in the QPO frequencies observed for GRO~J1655--40 and 4U~1630--47 may be 
related to a gradual increase in $r_{t}$ or a gradual drop in the mass 
accretion rate (or both).

In studies of the X-ray power spectra of BHC and neutron star X-ray binaries,
Wijnands \& van der Klis~(1999) find a correlation between the frequency of 
QPOs between 0.2 and 67~Hz and the break frequency of the continuum component 
(described as a flat-top component in this paper).  Such a correlation is 
interesting since it suggests that there is a physical property of the system 
that sets both time scales and that the physical property does not depend on the 
different properties of BHCs and neutron stars.  While 4U~1630--47 was in its hard 
state, the break frequency gradually decreased from $3.33\pm 0.36$~Hz to 
$0.48\pm 0.03$~Hz between observations 41 and 46-48 as the QPO frequency dropped 
from 3.4~Hz to 0.23~Hz (cf. see Tables~\ref{tab:powercont} and \ref{tab:shiftqpo}).  
As for the other sources included in the Wijnands \& van der Klis~(1999) sample, 
4U~1630--47 exhibits a correlation between QPO frequency and break frequency.  
However, for 4U~1630--47, the QPO frequency is below or consistent with the 
break frequency, while in other sources the QPO frequency is above the break
frequency.

\subsection{Emission Properties During the Flare}

\begin{figure}
\plotone{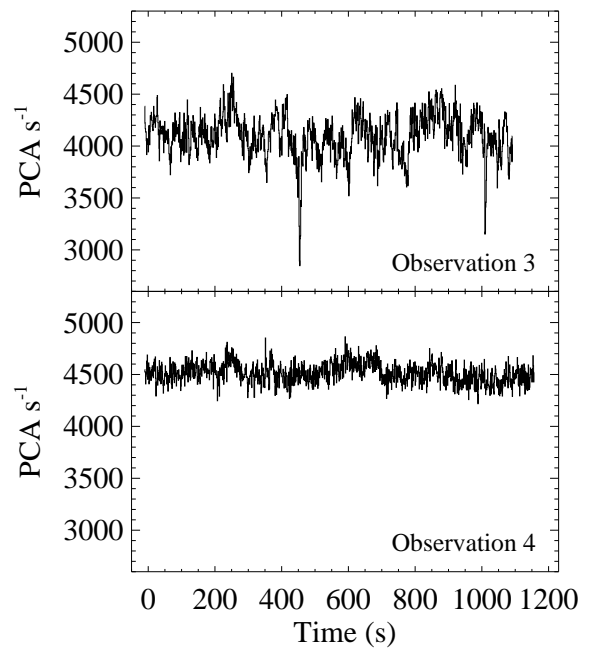}
\caption{PCA light curves for the flare observations (observations 3
and 4).  The 2-21~keV X-rays are binned in 1~s intervals, and 
background has not been subtracted.  The figure shows that the source 
variability was much higher during observation 3 than during 
observation 4.\label{fig:flare}}
\end{figure}

Figure~\ref{fig:flare} shows the 2-60~keV PCA light curves for the two observations 
made during the flare which occurred around MJD 50880 (observations 3 and 4).
For observation 3, short (about 4~s) X-ray dips are observed.  We have 
examined the light curves for all 51 observations and find that X-ray dips
are only observed for observation 3.  However, 4U~1630--47 observations
made by another group show that short X-ray dips were observed earlier in
the outburst (\cite{dieters99}).  In addition to the dips, Figure~\ref{fig:flare}
shows that the level of variability is much higher for observation 3
than for observation 4.  Table~\ref{tab:powercont} details the differences 
between the power spectra for these two observations.  For observation 3, 
the flat-top and power-law RMS amplitudes are 3.55\% and 4.36\%, respectively, 
while, for observation 4, the flat-top and power-law RMS amplitudes are 1.83\% 
and 1.10\%, which are even lower than most of the nearby non-flare observations.  
Also, QPOs are observed for observation 3 but not for observation 4.  The timing 
differences between these two observations are especially remarkable because 
the energy spectra for observations 3 and 4 are nearly identical 
(cf. Table~\ref{tab:energy}).  

The asymmetry of the low frequency QPO peak for observation 3 is similar to
QPOs observed for GS~1124--68 (\cite{belloni97}) and XTE~J1748--288 
(Revnivtsev et al.~1999).  For these two sources and for 4U~1630--47, the 
asymmetric shape of the QPO can be modeled using two Lorentzians, suggesting 
that the asymmetry may be due to a shift in the QPO centroid during the 
observation.  Revnivtsev et al.~(1999) find that some properties of the 
XTE~J1748--288 power spectra are consistent with this picture.
The 4U~1630--47 timing properties during observation 3 are not consistent 
with a gradual shift in the QPO centroid during the observation since the 
5.4~Hz and 6.2~Hz Lorentzians are present in both segments of the observation
(cf. Table~\ref{tab:obs3}).  The stability of the QPO shape may indicate that 
the asymmetric peak is caused by an intrinsic property of the QPO mechanism.
However, for observations of 4U~1630--47 containing dips, Dieters et al.~(1999) 
find that the frequencies of some QPOs are lower within the dips than outside 
the dips.  For our observation 3, it is possible that frequency changes during 
the dips (cf. Figure~\ref{fig:flare}) cause the QPO profile to be asymmetric.

\section{Summary and Conclusions}

We have analyzed data from 51 \it RXTE \rm observations of 4U~1630--47 during 
the decay of its 1998 outburst to study the evolution of its spectral and 
timing properties.  During the decay, the X-ray flux dropped exponentially 
with an e-folding time of about 14.4~d, which is short compared to most other 
BHC X-ray transients.  The e-folding time was nearly the same (14.9~d) for the 
decay of the 1996 outburst, which may indicate that this time scale is set by 
some property of the system that does not change between outbursts.  For the 
1998 outburst, the decay was interrupted by a secondary maximum, which is 
commonly observed for BHC X-ray transients.

Our analysis of the 4U~1630--47 power spectra indicates that 0.2~Hz to 11~Hz 
QPOs with RMS amplitudes between 2\% and 9\% occurred during the observations.  
During one of our early observations, when the source was relatively bright, 
a QPO occurred near 6~Hz with a profile that cannot be described by a single 
Lorentzian.  Similar asymmetric QPO peaks have been observed previously for 
GS~1124--68 (\cite{belloni97}) and XTE~J1748--288 (Revnivtsev et al.~1999).  
For all three sources (4U~1630--47, GS~1124--68 and XTE~J1748--288), the QPO 
is well-described by a combination of two Lorentzians.  

Near the end of the outburst, an abrupt change in the 4U~1630--47 spectral and 
timing properties occurred, and we interpret this change as evidence for a 
soft-to-hard state transition.  Our observations indicate that most of the changes 
in the emission properties, associated with the transition, occurred over a time 
period less than 2~d.  The timing properties changed after the transition with 
an increase in the continuum noise level and the appearance of a QPO.  A 3.4~Hz 
QPO appeared immediately after the transition, and, in subsequent observations, 
the QPO frequency decreased gradually to about 0.2~Hz.  At the transition, the 
energy spectrum also changed with an abrupt drop in the soft component flux in 
the \it RXTE \rm band pass, which was probably due to a drop in the inner disk 
temperature.  A change in the power-law photon index from 2.3 to 1.8, also 
associated with the transition, occurred over a time period of 8~d.  Although 
many of these changes are typical of soft-to-hard state transitions, the QPO 
behavior and the short time scale for the transition are not part of the 
canonical picture for state transitions (\cite{v95};~\cite{nowak95};~\cite{ct96}).
Finally, we note that 4U~1630--47 exhibits interesting behavior (e.g., state 
changes and QPOs) below a flux level of $10^{-9}$~erg~cm$^{-2}$~s$^{-1}$, 
indicating that observing programs for X-ray transients should be designed to 
follow these sources to low flux levels.

\acknowledgements

The authors would like to thank J.H. Swank for approving observations of 
4U~1630--47 at low flux levels, S. Dieters for providing results from 
\it BeppoSAX \rm observations prior to publication and an anonymous 
referee whose comments led to an improved paper.  We acknowledge partial 
support from NASA grants NAG5-4633, NAG5-4416 and NAG5-7347.


\begin{deluxetable}{lccc}
\footnotesize
\tablecaption{RXTE Observations of 4U~1630--47 \label{tab:obs}}
\tablewidth{0pt}
\tablehead{\colhead{Observation} & \colhead{MJD\tablenotemark{a}} & \colhead{Integration Time (s)} & \colhead{PCA count rate\tablenotemark{b}~(s$^{-1}$)}}
\startdata
1 & 50876.9325 & 1824 & 3158\nl
2 & 50877.1978 & 1712 & 3135\nl
3 & 50880.3884 & 1136 & 4005\nl
4 & 50881.3865 & 1232 & 4420\nl
5 & 50883.9011 & 7248 & 2453\nl
6 & 50885.7305 & 2832 & 2108\nl
7 & 50885.8687 & 6432 & 2051\nl
8 & 50887.8010 & 9920 & 1826\nl
9 & 50891.7337 & 9152 & 1400\nl
10 & 50892.6314 & 4816 & 1313\nl
11 & 50893.8026 & 9456 & 1212\nl
12 & 50895.7370 & 9424 & 992\nl
13 & 50899.0084 & 1200 & 822\nl
14 & 50900.6819 & 912 & 736\nl
15 & 50904.2146 & 992 & 626\nl
16 & 50904.6203 & 9328 & 621\nl
17 & 50906.5668 & 7056 & 587\nl
18 & 50907.5942 & 9920 & 568\nl
19 & 50909.5107 & 11120 & 553\nl
20 & 50911.3969 & 9920 & 559\nl
21 & 50913.3997 & 11040 & 547\nl
22 & 50921.5827 & 576 & 599\nl
23 & 50923.6826 & 9824 & 587\nl
24 & 50924.8245 & 10128 & 634\nl
25 & 50925.8665 & 1584 & 653\nl
26 & 50926.6551 & 400 & 692\nl
27 & 50927.7248 & 512 & 714\nl
28 & 50928.5927 & 416 & 723\nl
29 & 50929.6599 & 560 & 741\nl
30 & 50930.6609 & 608 & 772\nl
31 & 50931.8622 & 544 & 801\nl
32 & 50932.7974 & 800 & 825\nl
33 & 50935.1336 & 1136 & 854\nl
34 & 50936.1995 & 1024 & 873\nl
35 & 50937.6251 & 512 & 847\nl
36 & 50939.0651 & 736 & 799\nl
37 & 50942.0190 & 672 & 758\nl
38 & 50945.8624 & 352 & 571\nl
39 & 50947.8040 & 960 & 467\nl
40 & 50949.7373 & 864 & 404\nl
41 & 50951.6677 & 1328 & 397\nl
42 & 50952.4870 & 880 & 361\nl
43 & 50953.4910 & 1408 & 307\nl
44 & 50956.9571 & 1440 & 201\nl
45 & 50958.9578 & 1312 & 208\nl
46 & 50961.0918 & 912 & 194\nl
47 & 50962.9571 & 1440 & 240\nl
48 & 50965.9570 & 1616 & 207\nl
49 & 50968.2637 & 624 & 193\nl
50 & 50970.3910 & 720 & 145\nl
51 & 50972.1303 & 864 & 203\nl
\tablenotetext{a}{Modified Julian Date (MJD = JD--2400000.5) at the midpoint of the observation.}
\tablenotetext{b}{2.5-20~keV count rate for 5 PCUs (all 3 layers) after background subtraction.}
\enddata
\end{deluxetable}

\begin{deluxetable}{lcccccc}
\footnotesize
\tablecaption{Power Spectra: Continuum Parameters\tablenotemark{a}\label{tab:powercont}}
\tablewidth{0pt}
\tablehead{ & \multicolumn{3}{c}{Flat-top} & \multicolumn{2}{c}{Power-law} & }
\startdata
Obs. & RMS\tablenotemark{b}~(\%) & $\nu_{break}$~(Hz) & $\alpha$ & RMS\tablenotemark{b}~(\%) & $\alpha$ & $\chi^{2}/\nu$\nl
\hline
1-2\tablenotemark{e} & $3.24\pm 0.21$ & $0.49\pm 0.05$ & $-0.96\pm 0.04$ & $1.37\pm 0.66$ 
	& $-1.56\pm 0.19$ & 471/438\nl
3\tablenotemark{e} & $3.55\pm 0.21$ & $2.72\pm 0.28$ & $-1.46\pm 0.15$ & $4.36\pm 0.72$ 
	& $-1.70\pm 0.07$ & 552/432\nl
4 & $1.83\pm 0.33$ & $1.52\pm 0.66$ & $-0.73\pm 0.11$ & $1.10\pm 0.56$ 
	& $-1.52\pm 0.24$ & 495/441\nl
5 & $2.04\pm 0.28$ & $2.54\pm 0.50$ & $-1.23\pm 0.28$ & $1.98\pm 0.32$ 
	& $-0.97\pm 0.08$ & 418/441\nl
6\tablenotemark{e} & $2.71\pm 0.27$ & $1.42\pm 0.14$ & $-1.90\pm 0.33$ & $1.60\pm 0.48$ 
	& $-1.05\pm 0.16$ & 453/438\nl
7\tablenotemark{e} & $2.49\pm 0.18$ & $1.72\pm 0.10$ & $-2.87\pm 0.54$ & $2.47\pm 0.23$ 
	& $-0.89\pm 0.05$ & 405/438\nl
8\tablenotemark{e} & $2.48\pm 0.18$ & $1.40\pm 0.09$ & $-2.24\pm 0.38$ & $2.38\pm 0.15$ 
	& $-0.81\pm 0.04$ & 470/438\nl
9-10 & $2.67\pm 0.22$ & $1.60\pm 0.14$ & $-1.69\pm 0.24$ & $2.06\pm 0.25$ 
	& $-0.75\pm 0.07$ & 431/441\nl
11-12 & $3.52\pm 0.16$ & $0.56\pm 0.04$ & $-1.05\pm 0.04$ & $<1.0$\tablenotemark{c}
	& $-1.0$\tablenotemark{d} & 518/443\nl
13-24 & $3.65\pm 0.12$ & $0.76\pm 0.04$ & $-1.33\pm 0.06$ & $<0.7$\tablenotemark{c} 
	& $-1.0$\tablenotemark{d} & 421/443\nl
25-40 & $3.13\pm 0.27$ & $1.57\pm 0.14$ & $-2.82\pm 0.71$ & $<1.0$\tablenotemark{c}
	& $-1.0$\tablenotemark{d} & 452/443\nl
41\tablenotemark{e} & $10.2\pm 0.6$ & $3.33\pm 0.36$ & $-1.73\pm 0.24$ 
	& $<2.5$\tablenotemark{c} & $-1.0$\tablenotemark{d} & 480/440\nl
42\tablenotemark{e} & $11.3\pm 0.8$ & $2.81\pm 0.33$ & $-1.60\pm 0.21$ 
	& $<3.6$\tablenotemark{c} & $-1.0$\tablenotemark{d} & 415/440\nl
43\tablenotemark{e} & $13.6\pm 0.7$ & $1.97\pm 0.12$ & $-1.87\pm 0.16$ 
	& $<3.3$\tablenotemark{c} & $-1.0$\tablenotemark{d} & 479/440\nl
44\tablenotemark{e} & $15.9\pm 1.1$ & $0.83\pm 0.07$ & $-1.45\pm 0.10$ 
	& $<5.2$\tablenotemark{c} & $-1.0$\tablenotemark{d} & 550/440\nl
45\tablenotemark{e} & $16.1\pm 1.5$ & $0.55\pm 0.06$ & $-1.31\pm 0.08$ 
	& $<2.0$\tablenotemark{c} & $-1.0$\tablenotemark{d} & 491/440\nl
46-48\tablenotemark{e} & $17.3\pm 0.8$ & $0.48\pm 0.03$ & $-1.33\pm 0.05$ 
	& $<5.4$\tablenotemark{c} & $-1.0$\tablenotemark{d} & 453/440\nl
49\tablenotemark{e} & $14.9\pm 2.2$ & $0.26\pm 0.05$ & $-1.45\pm 0.12$ 
	& $<9.0$\tablenotemark{c} & $-1.0$\tablenotemark{d} & 550/440\nl
50-51 & $13.8\pm 1.1$ & $0.19\pm 0.02$ & $-1.28\pm 0.08$ & $<8.3$\tablenotemark{c}
	& $-1.0$\tablenotemark{d} & 494/440\nl
\tablenotetext{a}{The errors correspond to $\Delta \chi^{2} = 1.0$ (68\% confidence).}
\tablenotetext{b}{0.01-10~Hz RMS amplitudes.}
\tablenotetext{c}{90\% confidence upper limit.}
\tablenotetext{d}{Power-law index fixed to this value.}
\tablenotetext{e}{A Lorentzian is included in the model.  For observation 3, three
Lorentzians are included in the model as described in the text.}
\enddata
\end{deluxetable}

\begin{deluxetable}{lccccc}
\footnotesize
\tablecaption{QPO Parameters for Observation 3\tablenotemark{a}\label{tab:obs3}}
\tablewidth{0pt}

\tablehead{ & \multicolumn{3}{c}{All of Observation 3} & 
\multicolumn{1}{c}{Time Segment 1\tablenotemark{b}} & 
\multicolumn{1}{c}{Time Segment 2\tablenotemark{c}}}
\startdata
 & 1 Lorentzian & 2 Lorentzians & 3 Lorentzians & 3 Lorentzians & 3 Lorentzians\nl
\hline
$\nu_{1}$ (Hz) & $5.74\pm 0.03$ & $5.42\pm 0.02$ & $5.43\pm 0.02$ & $5.49\pm 0.02$ & $5.35\pm 0.03$\nl
FWHM$_{1}$ (Hz) & $1.06\pm 0.07$ & $0.37\pm 0.06$ & $0.41\pm 0.06$ & $0.38\pm 0.05$ & $0.49\pm 0.10$\nl
RMS$_{1}$ (\%) & $4.03\pm 0.10$ & $2.76\pm 0.19$ & $2.89\pm 0.18$ & $2.95\pm 0.20$ & $2.88\pm 0.25$\nl
\hline
$\nu_{2}$ (Hz) & --- & $6.17\pm 0.05$ & $6.19\pm 0.04$ & $6.17\pm 0.06$ & $6.25\pm 0.06$\nl
FWHM$_{2}$ (Hz) & --- & $0.77\pm 0.13$ & $0.78\pm 0.12$ & $0.71\pm 0.16$ & $0.85\pm 0.17$\nl
RMS$_{2}$ (\%) & --- & $2.85\pm 0.22$ & $2.85\pm 0.21$ & $2.60\pm 0.25$ & $3.09\pm 0.27$\nl
\hline
$\nu_{3}$ (Hz) & --- & --- & $10.79\pm 0.14$ & $11.00\pm 0.19$ & $10.57\pm 0.18$\nl
FWHM$_{3}$ (Hz) & --- & --- & $1.47\pm 0.45$ & $2.00\pm 0.63$ & $1.56\pm 0.53$\nl
RMS$_{3}$ (\%) & --- & --- & $1.85\pm 0.20$ & $2.11\pm 0.24$ & $1.95\pm 0.26$\nl
\hline
$\chi^{2}/\nu$ & 685/438 & 608/435 & 552/432 & 570/432 & 480/432\nl
\tablenotetext{a}{The errors correspond to $\Delta \chi^{2} = 1.0$ (68\% confidence).}
\tablenotetext{b}{First 576 seconds of Observation 3.}
\tablenotetext{c}{Last 512 seconds of Observation 3.}
\enddata
\end{deluxetable}

\begin{deluxetable}{lccc}
\footnotesize
\tablecaption{QPO Parameters for Observations 41-49\tablenotemark{a}\label{tab:shiftqpo}}
\tablewidth{0pt}
\tablehead{\colhead{Observation} & \colhead{Frequency (Hz)} & \colhead{FWHM (Hz)} & \colhead{RMS (\%)}}
\startdata
41 & $3.390\pm 0.008$ & $0.14\pm 0.02$ & $7.30\pm 0.33$\nl
42 & $2.613\pm 0.012$ & $0.17\pm 0.03$ & $8.46\pm 0.47$\nl
43 & $1.351\pm 0.012$ & $0.22\pm 0.04$ & $8.68\pm 0.51$\nl
44 & $0.430\pm 0.006$ & $0.07\pm 0.02$ & $<8.2$\tablenotemark{b}\nl
45 & $0.365\pm 0.011$ & $0.11\pm 0.04$ & $<9.6$\tablenotemark{b}\nl
46-48 & $0.228\pm 0.003$ & $0.046\pm 0.010$ & $6.55\pm 0.51$\nl
49 & $0.182\pm 0.005$ & $0.043\pm 0.012$ & $<10.1$\tablenotemark{b}\nl
\tablenotetext{a}{The errors correspond to $\Delta \chi^{2} = 1.0$ (68\% confidence).}
\tablenotetext{b}{90\% confidence upper limit.}
\enddata
\end{deluxetable}

\begin{deluxetable}{lccccccccccc}
\scriptsize
\tablecaption{Energy Spectrum Fit Parameters\tablenotemark{a,b}\label{tab:energy}}
\tablewidth{0pt}
\tablehead{ & \multicolumn{2}{c}{Power-law} & \multicolumn{2}{c}{Disk-blackbody} & 
\multicolumn{3}{c}{Narrow Emission Line} & \multicolumn{2}{c}{Broad Absorption Edge}}
\startdata
Obs. & $\Gamma$ & $F_{PL}$\tablenotemark{c} &
$kT_{in}$ (keV) & $F_{DBB}$\tablenotemark{d} & 
$E_{line}$ (keV) & $N_{line}$\tablenotemark{e} & Signif.~(\%) & 
$E_{edge}$ & $\tau_{\rm{Fe}}$ & $\chi^{2}/\nu$\nl
\hline
1 & $2.498\pm 0.019$ & 7.42 & 0.955 & 7.95 & 
6.60\tablenotemark{g} & $<2.6$\tablenotemark{f} & 72.8 & 
$8.65\pm 0.16$ & $0.84\pm 0.09$ & 80/102\nl
2 & $2.389\pm 0.017$ & 7.34 & 0.917 & 8.66 & 
6.68\tablenotemark{g} & $<3.1$\tablenotemark{f} & 78.8 & 
$9.05\pm 0.13$ & $0.94\pm 0.08$ & 135/102\nl
3 & $2.603\pm 0.015$ & 10.6 & 1.577 & $<2.79$ & 
6.41\tablenotemark{g} & $<2.4$\tablenotemark{f} & 52.3 & 
$8.71\pm 0.34$ & $0.43\pm 0.10$ & 96/102\nl
4 & $2.557\pm 0.012$ & 10.3 & 1.573 & $<5.08$ & 
6.79\tablenotemark{g} & $<1.9$\tablenotemark{f} & 67.0 & 
$9.15\pm 0.32$ & $0.35\pm 0.08$ & 107/102\nl
5 & $2.500\pm 0.021$ & 3.70 & 1.049 & 11.5 & 
6.63\tablenotemark{g} & $<2.3$\tablenotemark{f} & 87.3 & 
$9.00\pm 0.09$ & $1.50\pm 0.10$ & 90/102\nl
6 & $2.402\pm 0.023$ & 3.99 & 0.933 & 9.62 & 
$6.70\pm 0.07$ & $1.9\pm 0.5$ & 94.7 & 
$8.93\pm 0.10$ & $1.42\pm 0.10$ & 104/102\nl
7 & $2.457\pm 0.016$ & 4.09 & 0.905 & 9.14 & 
$6.61\pm 0.08$ & $1.7\pm 0.5$ & 93.2 & 
$8.77\pm 0.09$ & $1.36\pm 0.08$ & 92/102\nl
8 & $2.453\pm 0.016$ & 3.50 & 0.914 & 8.61 & 
$6.63\pm 0.07$ & $1.6\pm 0.4$ & 97.4 & 
$8.76\pm 0.08$ & $1.47\pm 0.08$ & 78/102\nl
9 & $2.412\pm 0.017$ & 2.61 & 0.861 & 7.94 & 
$6.60\pm 0.07$ & $1.2\pm 0.3$ & 93.6 & 
$8.55\pm 0.08$ & $1.55\pm 0.08$ & 90/102\nl
10 & $2.396\pm 0.023$ & 2.38 & 0.857 & 7.96 & 
$6.65\pm 0.07$ & $1.2\pm 0.3$ & 94.4 & 
$8.61\pm 0.09$ & $1.78\pm 0.10$ & 93/102\nl
11 & $2.444\pm 0.020$ & 2.14 & 0.847 & 7.83 & 
$6.52\pm 0.06$ & $1.3\pm 0.3$ & 97.4 & 
$8.59\pm 0.07$ & $1.84\pm 0.09$ & 104/102\nl
12 & $2.276\pm 0.015$ & 2.14 & 0.714 & 7.13 & 
$6.43\pm 0.07$ & $1.0\pm 0.3$ & 97.7 & 
$8.37\pm 0.08$ & $1.46\pm 0.07$ & 78/102\nl
13 & $2.265\pm 0.040$ & 1.67 & 0.733 & 6.28 & 
6.49\tablenotemark{g} & $<1.5$\tablenotemark{f} & 88.2 & 
$8.57\pm 0.15$ & $1.56\pm 0.17$ & 97/102\nl
14 & $2.265\pm 0.042$ & 1.63 & 0.666 & 6.30 & 
6.25\tablenotemark{g} & $<1.3$\tablenotemark{f} & 78.6 & 
$8.22\pm 0.18$ & $1.43\pm 0.19$ & 83/102\nl
15 & $2.264\pm 0.050$ & 1.37 & 0.664 & 5.45 & 
6.32\tablenotemark{g} & $<1.2$\tablenotemark{f} & 85.1 & 
$8.37\pm 0.18$ & $1.49\pm 0.22$ & 87/102\nl
16 & $2.248\pm 0.017$ & 1.43 & 0.636 & 5.58 & 
6.47\tablenotemark{g} & $<0.8$\tablenotemark{f} & 88.8 & 
$8.08\pm 0.10$ & $1.40\pm 0.09$ & 64/102\nl
17 & $2.122\pm 0.017$ & 1.41 & 0.605 & 5.22 & 
$6.49\pm 0.07$ & $0.6\pm 0.2$ & 92.9 & 
$8.23\pm 0.11$ & $1.24\pm 0.09$ & 77/102\nl
18 & $2.136\pm 0.016$ & 1.39 & 0.595 & 5.09 & 
$6.49\pm 0.06$ & $0.7\pm 0.2$ & 96.3 & 
$8.44\pm 0.09$ & $1.30\pm 0.08$ & 86/102\nl
19 & $2.074\pm 0.015$ & 1.28 & 0.643 & 4.95 & 
$6.50\pm 0.05$ & $0.7\pm 0.2$ & 96.5 & 
$8.31\pm 0.09$ & $1.49\pm 0.08$ & 100/102\nl
20 & $2.264\pm 0.021$ & 1.14 & 0.701 & 4.81 & 
$6.54\pm 0.06$ & $0.7\pm 0.2$ & 98.2 & 
$8.31\pm 0.09$ & $1.61\pm 0.10$ & 73/102\nl
21 & $2.330\pm 0.024$ & 1.05 & 0.718 & 4.93 & 
$6.50\pm 0.06$ & $0.7\pm 0.2$ & 98.7 & 
$8.45\pm 0.09$ & $1.75\pm 0.10$ & 87/102\nl
22 & $2.064\pm 0.050$ & 1.40 & 0.633 & 4.97 & 
6.38\tablenotemark{g} & $<1.0$\tablenotemark{f} & 63.0 & 
$8.20\pm 0.24$ & $1.34\pm 0.24$ & 101/102\nl
23 & $2.202\pm 0.021$ & 1.24 & 0.701 & 4.20 & 
$6.56\pm 0.05$ & $0.8\pm 0.2$ & 99.5 & 
$8.61\pm 0.08$ & $1.75\pm 0.09$ & 90/102\nl
24 & $2.185\pm 0.017$ & 1.43 & 0.681 & 4.68 & 
6.49\tablenotemark{g} & $<0.8$\tablenotemark{f} & 87.3 & 
$8.13\pm 0.10$ & $1.35\pm 0.09$ & 90/102\nl
25 & $2.180\pm 0.036$ & 1.45 & 0.693 & 4.83 & 
6.59\tablenotemark{g} & $<1.1$\tablenotemark{f} & 82.6 & 
$8.32\pm 0.18$ & $1.30\pm 0.16$ & 88/102\nl
26 & $2.158\pm 0.066$ & 1.46 & 0.723 & 4.69 & 
6.31\tablenotemark{g} & $<1.6$\tablenotemark{f} & 73.5 & 
$8.18\pm 0.33$ & $1.22\pm 0.29$ & 84/102\nl
27 & $2.269\pm 0.065$ & 1.51 & 0.731 & 5.10 & 
6.48\tablenotemark{g} & $<1.2$\tablenotemark{f} & 60.4 & 
$8.14\pm 0.27$ & $1.42\pm 0.28$ & 94/102\nl
28 & $2.466\pm 0.068$ & 1.64 & 0.706 & 4.74 & 
6.32\tablenotemark{g} & $<1.1$\tablenotemark{f} & 54.5 & 
$7.95\pm 0.42$ & $0.99\pm 0.29$ & 93/102\nl
29 & $2.320\pm 0.071$ & 1.52 & 0.760 & 4.89 & 
6.77\tablenotemark{g} & $<1.4$\tablenotemark{f} & 66.4 & 
$8.71\pm 0.28$ & $1.27\pm 0.28$ & 92/102\nl
30 & $2.364\pm 0.064$ & 1.58 & 0.762 & 5.25 & 
6.56\tablenotemark{g} & $<1.4$\tablenotemark{f} & 72.0 & 
$8.46\pm 0.19$ & $1.68\pm 0.25$ & 104/102\nl
31 & $2.370\pm 0.060$ & 1.70 & 0.757 & 5.08 & 
6.33\tablenotemark{g} & $<1.3$\tablenotemark{f} & 65.3 & 
$8.31\pm 0.26$ & $1.29\pm 0.25$ & 88/102\nl
32 & $2.268\pm 0.048$ & 1.73 & 0.750 & 5.40 & 
$6.35\pm 0.08$ & $1.3\pm 0.4$ & 90.1 & 
$8.53\pm 0.19$ & $1.44\pm 0.20$ & 73/102\nl
33 & $2.348\pm 0.042$ & 1.73 & 0.773 & 5.68 & 
6.55\tablenotemark{g} & $<1.3$\tablenotemark{f} & 70.7 & 
$8.09\pm 0.18$ & $1.52\pm 0.19$ & 81/102\nl
35 & $2.367\pm 0.067$ & 1.70 & 0.764 & 5.99 & 
6.62\tablenotemark{g} & $<1.0$\tablenotemark{f} & 53.2 & 
$8.24\pm 0.18$ & $2.10\pm 0.28$ & 111/102\nl
36 & $2.356\pm 0.052$ & 1.67 & 0.738 & 5.75 & 
6.39\tablenotemark{g} & $<1.2$\tablenotemark{f} & 63.2 & 
$8.40\pm 0.18$ & $1.62\pm 0.22$ & 89/102\nl
37 & $2.422\pm 0.068$ & 1.40 & 0.758 & 6.24 & 
6.36\tablenotemark{g} & $<1.3$\tablenotemark{f} & 73.3 & 
$8.37\pm 0.19$ & $1.89\pm 0.19$ & 85/102\nl
38 & $2.317\pm 0.092$ & 1.12 & 0.711 & 5.01 & 
6.65\tablenotemark{g} & $<0.8$\tablenotemark{f} & 51.0 & 
$8.38\pm 0.39$ & $1.28\pm 0.39$ & 81/102\nl
40 & $2.072\pm 0.058$ & 0.90 & 0.601 & 4.78 & 
$6.33\pm 0.07$ & $0.8\pm 0.3$ & 99.96 & 
$8.47\pm 0.18$ & $1.73\pm 0.27$ & 75/102\nl
41 & $1.916\pm 0.027$ & 1.09 & $<0.461$ & $<4.69$ & 
6.45\tablenotemark{g} & $<0.8$\tablenotemark{f} & 80.1 & 
$8.06\pm 0.22$ & $0.92\pm 0.18$ & 107/102\nl
42 & $1.858\pm 0.032$ & 1.00 & $<0.466$ & $<3.57$ & 
6.43\tablenotemark{g} & $<0.6$\tablenotemark{f} & 58.4 & 
$7.74\pm 0.25$ & $1.08\pm 0.22$ & 63/102\nl
43 & $1.798\pm 0.027$ & 0.85 & $<0.487$ & $<2.42$ & 
6.51\tablenotemark{g} & $<0.6$\tablenotemark{f} & 71.8 & 
$7.49\pm 0.27$ & $0.91\pm 0.19$ & 98/102\nl
44 & $1.793\pm 0.037$ & 0.56 & $<0.459$ & $<2.19$ & 
6.44\tablenotemark{g} & $<0.4$\tablenotemark{f} & 61.6 & 
$7.54\pm 0.29$ & $1.05\pm 0.26$ & 74/102\nl
45 & $1.675\pm 0.031$ & 0.58 & $<0.486$ & $<1.68$ & 
6.46\tablenotemark{g} & $<0.6$\tablenotemark{f} & 85.6 & 
7.1\tablenotemark{g} & $1.36\pm 0.27$ & 89/103\nl
46 & $1.735\pm 0.038$ & 0.55 & $<0.451$ & $<1.83$ & 
6.54\tablenotemark{g} & $<0.6$\tablenotemark{f} & 73.4 & 
7.1\tablenotemark{g} & $1.19\pm 0.33$ & 78/103\nl
47 & $1.584\pm 0.024$ & 0.67 & $<0.452$ & $<1.70$ & 
$6.47\pm 0.07$ & $0.5\pm 0.2$ & 93.0 & 
7.1\tablenotemark{g} & $0.66\pm 0.22$ & 74/103\nl
49 & $1.640\pm 0.055$ & 0.54 & $<0.493$ & $<1.27$ & 
6.34\tablenotemark{g} & $<0.7$\tablenotemark{f} & 68.3 & 
$7.56\pm 0.32$ & $1.45\pm 0.37$ & 83/102\nl
50 & $1.536\pm 0.040$ & 0.50 & $<0.495$ & $<1.21$ & 
6.50\tablenotemark{g} & $<0.7$\tablenotemark{f} & 74.5 & 
7.1\tablenotemark{g} & $0.98\pm 0.36$ & 99/103\nl
51 & $1.605\pm 0.031$ & 0.57 & $<0.365$ & $<3.20$ & 
6.40\tablenotemark{g} & $<0.6$\tablenotemark{f} & 72.7 & 
7.1\tablenotemark{g} & $0.78\pm 0.28$ & 94/103\nl
44-51 & $1.657\pm 0.012$ & 0.58 & $<0.455$ & $<1.83$ & 
$6.46\pm 0.04$ & $0.4\pm 0.1$ & 99.93 & 
7.1\tablenotemark{g} & $1.08\pm 0.11$ & 98/103\nl
\tablenotetext{a}{Column density fixed to $9.45\times 10^{22}$~cm$^{-2}$.}
\tablenotetext{b}{The errors correspond to $\Delta \chi^{2} = 1.0$ (68\% confidence).}
\tablenotetext{c}{2.5-20~keV unabsorbed flux in units of
10$^{-9}$~erg~cm$^{-2}$~s$^{-1}$.}
\tablenotetext{d}{Bolometric flux in units of 10$^{-9}$~erg~cm$^{-2}$~s$^{-1}$.}
\tablenotetext{e}{Normalization in units of 10$^{-3}$~photons~cm$^{-2}$~s$^{-1}$.}
\tablenotetext{f}{90\% confidence upper limit.}
\tablenotetext{g}{Fixed.}
\enddata
\end{deluxetable}

\appendix

\section{SGR~1627--41}

Soft $\gamma$-ray bursts were detected from a position near 4U~1630--47 on 
MJD 50979 (\cite{kouveliotou98}), 7~d after our last \it RXTE \rm observation.  
The soft $\gamma$-ray repeater, SGR~1627--41, was observed with \it RXTE \rm on 
MJD 50990, and a 0.15~Hz QPO was detected during the observation (\cite{dieters98b}).  
Although the position of SGR~1627--41 is not consistent with the position of 
4U~1630--47 (\cite{hurley99}), the two sources are close enough so that they were 
both in the \it RXTE \rm field of view during the observation made on MJD 50990
and also during our observations, allowing for the possibility of source confusion. 
We inspected the \it RXTE \rm 0.125~s light curves for our 4U~1630--47 observations, 
and there is no evidence for activity (e.g., bursts) from SGR~1627--41.  An \it RXTE \rm 
scanning observation made on 1998 June 21 (MJD 50985) and \it BeppoSAX \rm observations 
made on 1998 August 7 (MJD 51032) and 1998 September 16 (MJD 51072) provide information 
about possible source confusion.  The scanning observation indicates that 4U~1630--47 
was much brighter than SGR~1627--41 on June 21.  Below, we present an analysis of the 
data from the scanning observation.  4U~1630--47 was also much brighter than 
SGR~1627--41 during the \it BeppoSAX \rm observations.  On August 7 and September 16, the 
2-10~keV unabsorbed flux for 4U~1630--47 was 30 to 40 times higher than for SGR~1627--41 
(\cite{woods99};~\cite{dieters99}).  It is likely that 4U~1630--47 also dominated the 
flux detected during the June 26 \it RXTE \rm observation and that it is responsible 
for the 0.15~Hz QPO.  Given the low persistent flux detected for SGR~1627--41 by 
\it BeppoSAX\rm, $6.7\times 10^{-12}$~erg~cm$^{-2}$~s$^{-1}$ unabsorbed in the 2-10~keV 
band (\cite{woods99}), it seems very unlikely that this source could be bright enough to 
produce the QPOs observed during our observations.

After soft $\gamma$-ray bursts were detected from SGR~1627--41
by BATSE (Burst and Transient Source Experiment) on 1998 June 15 
(\cite{kouveliotou98}), \it RXTE \rm scanning observations were 
made to locate a source of persistent X-ray emission related to 
soft $\gamma$-ray repeater (SGR).  When the scans were made, the 
position of SGR~1627--41 was restricted to the IPN (3rd 
Interplanetary Network) annulus reported in Hurley et al.~(1998a),
which is consistent with the position of the supernova remnant
G337.0-0.1.  \it RXTE \rm scans were made along the IPN annulus 
on 1998 June 19 and nearly perpendicular to the IPN annulus on 1998 
June 21.  Since other SGRs are associated with supernova remnants,
the perpendicular scan was centered on G337.0-0.1.  In the following 
months, the IPN position was improved (\cite{hurley98b}) and a
source of persistent X-ray emission related to the SGR was discovered
using \it BeppoSAX \rm (\cite{woods99}).  These observations restrict the 
SGR~1627--41 position to a 2$^{\prime}$ by 16$^{\prime\prime}$ region 
that is consistent with the position of G337.0-0.1, making an association 
between the two likely (\cite{hurley99}).

\setcounter{figure}{0}

\begin{figure}
\plotone{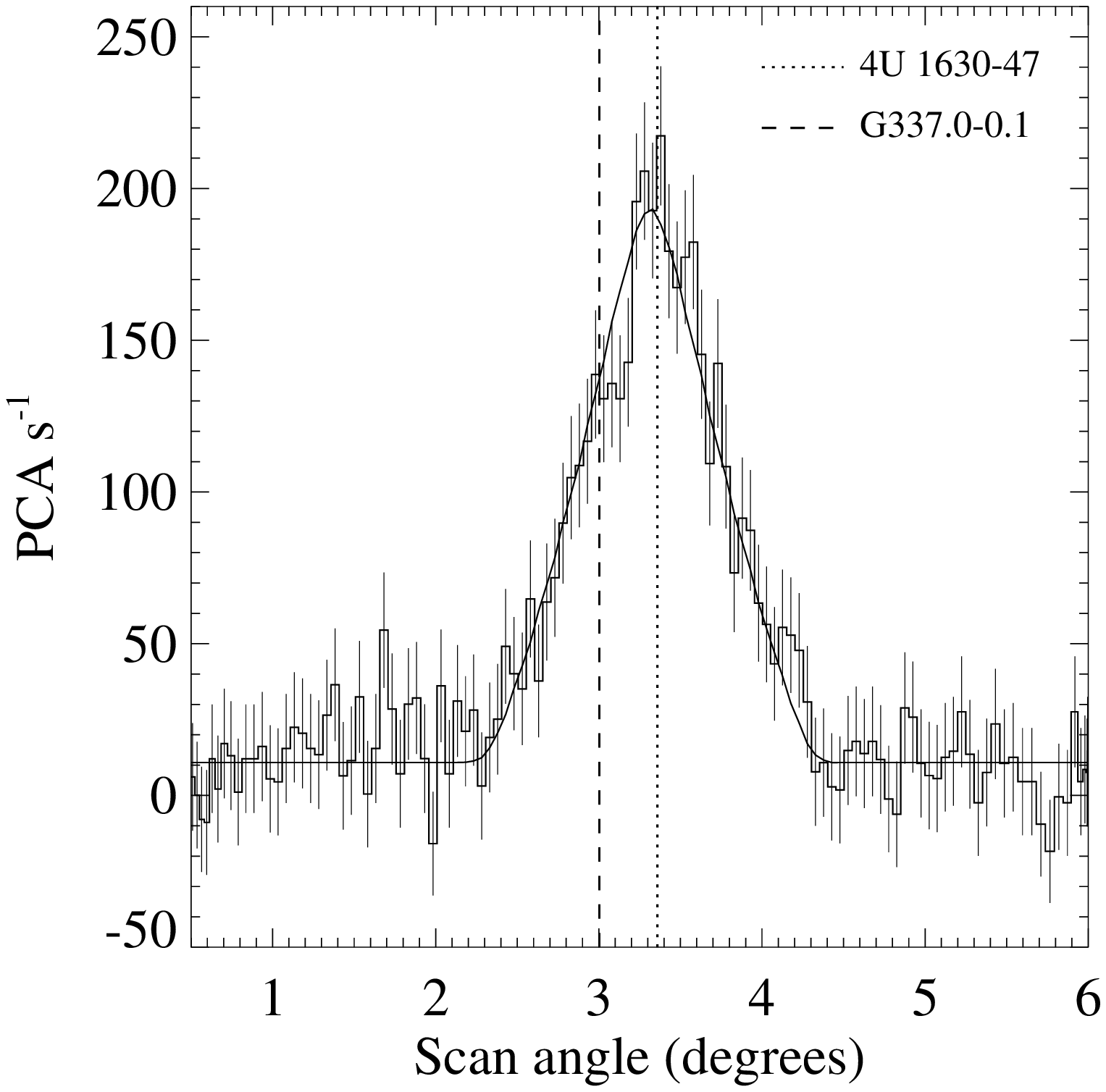}
\caption{The PCA light curve for the scan performed on June 21,
showing that X-ray emission was coming from a position consistent
with the 4U~1630--47 position (dotted line).  It is likely
that SGR~1627--41 is associated with the supernova remnant G337.0-0.1 
(\cite{hurley99}) and its position is also marked (dashed line).
\label{fig:scan}}
\end{figure}

We analyzed the \it RXTE \rm data from the June 21 scan to determine
if the persistent X-ray emission from SGR~1627--41 could have been bright 
enough to contaminate our \it RXTE \rm observations of 4U~1630--47.  
The linear scan passed through the positions of both G337.0-0.1 and 
4U~1630--47 for this purpose.  Figure~\ref{fig:scan} shows the background subtracted 
2-60~keV PCA count rate versus scan angle.  We fitted the light curve 
using a model consisting of a single point source and a constant count 
rate offset to account for small uncertainties in the background 
subtraction.  We used the 1996 June 5 PCA collimator response to
model the scan light curve produced by a point source.  A good fit is 
achieved ($\chi^{2}/\nu = 91/161$), indicating that the light curve 
is consistent with the presence of one source.  Figure~\ref{fig:scan} 
shows that the source position is consistent with 4U~1630--47 and not 
G337.0-0.1.  Also, the source amplitude is about 187~s$^{-1}$ (2-60~keV, 5 PCUs), 
which is close to the count rates reported for observations 44 to 51 
in Table~\ref{tab:obs}.  The \it RXTE \rm scan indicates that it is very unlikely 
that our 4U~1630--47 observations are significantly contaminated by 
emission from SGR~1627--41.

\end{document}